\documentstyle[12pt]{article}

\begin{document}
\newcommand{\be}{\begin{equation}}
\newcommand{\ee}{\end{equation}}
\newcommand{\bee}{\begin{ eqnarray }}
\newcommand{\eee}{\end{ eqnarray }}
\newcommand{\nn}{\nonumber \\ }


\title{ Chaos in  monopole sector \\  of  the Georgi - Glashow model.}

\author{
{\sc T. Dobrowolski and J. Szcz\c{e}sny}
\\ Institute of Physics and Informatics WSP, 
\\ Podchor\c{a}\.zych  2, 30-084 Cracow, Poland
}

\date{}
\maketitle

\begin{abstract}
A spherically symmetric excitations of the Polyakov - 't Hooft monopole  
are considered. In the framework of the geodesics deviation equation 
it is found that in the large mass Higgs sector a signature of chaos occurs.

\end{abstract}

\newpage

\section{ Introduction}

Because of extreme complication
of the evolution of nonlinear dynamical systems
 a great career  make methods originated in
Poincare section method \cite{Gucken}. There is also growing pressure on 
evaluating Lapunov exponents and Hausdorff dimension which
are  the most restrictive (global) characteristics of the  behavior
of the dynamical system \cite{Ott}. Simultaneously  there is also a tendency to
improve and find new local criterions of chaos which are much
simpler in application than the global one. Historically the first
criterion was based on the fact that in the neighbourhood of the
fixed points the dynamics is dominated by the linear terms \cite{Perko}.
This method in more formal way could be expressed as a 
condition (Toda criterion) on the Gauss curvature of the potential 
energy surface \cite{Toda}. 
The criterion based on the geodesics deviation equation 
in space equipped in Jacobi metric \cite{Arnold}
seems to be the best geometrically motivated  one.

On the other hand in particle physics \cite{Tze}, cosmology  \cite{Gal} and condensed
matter physics \cite{Zhang} there is (caused by growing number of applications)
tremendous interest in dynamics of the Yang - Mills fields.
On classical level, studies of this subject concentrate mainly on
seeking for new solutions and their excitations \cite{Dobro}. 

In this paper extending approach of the paper \cite{Ost} we apply local
geometric criterion of chaos to monopole sector of the Georgi - Glashow
model. Using the local criterion of chaos we investigate
the neighbourhood of the core of the Polyakov - 't Hooft  monopole.
We show that in the large mass Higgs sector  of the model, excitations
of the monopole behaves in chaotic way.
On the other hand for sufficiently  light Higgs fields
trajectories do not diverge.

In the next section we will remind basic facts about Georgi - Glashow 
model and monopole solutions. Then we will use 
the criterion of sensitive dependence on initial conditions
based on  geodesic deviation equation.
The suitable form of this criterion is prepared
in appendix. 
As a result we shall discover difference of the dynamics of the light 
and heavy Higgs mass sectors. Last section contains some remarks.  
    
\section{Spherical excitations of the core of the Polyakov - Thooft monopole.}

The Georgi - Glashow model is  $SO(3)$ gauge invariant system consisting
of a Higgs multiplet $\phi^a$ ($a=1,2,3$) transforming as a vector in the 
adjoint representation of the gauge group and the gauge fields
$A_{\mu} = {A^a}_{\mu}T^a$.

In the same representation of the gauge group generators take the 
form $(T^a)_{bc} = - i {f^a}_{bc}$, where $f^{abc} = \epsilon^{abc}$.
They also satisfy the relation $[T^a,T^b] = i f^{abc} T^c$.
The lagrangian of the model is given by
\be
{\cal {L}} = - {1 \over 4} {F^a}_{\mu\nu}{F^a}^{\mu\nu} + 
{1 \over 2} D_{\mu} \phi^a D^{\mu} \phi^a - V(\phi) ,
\ee
where
\be
F_{\mu\nu} = \partial_{\mu} A_{\nu} - \partial_{\nu} A_{\mu} + 
i e [A_{\mu}, A_{\nu}] 
\ee
are the field strengths, 
\be
D_{\mu} \phi^a = \partial_{\mu} \phi^a - e \epsilon^{abc} {A^b}_{\mu} \phi^c 
\ee
are covariant derivatives of the Higgs fields and
\be
V(\phi) = {1 \over 4} \lambda (\phi^2 - v^2)^2 
\ee
is a Higgs potential.
The Euler - Lagrange equations for lagrangian (1)
\be
( D_{\nu} F^{\mu\nu} )^a = - e \epsilon^{abc} \phi^b (D^{\mu} \phi)^c ,
\ee
\be
D_{\mu}D^{\mu} \phi^a = - \lambda \phi^a (\phi^2 - v^2) . 
\ee
The field strengths also satisfy the Bianchi identities.
Equations of motion have finite energy and time independent
solutions. Existence and stability of these solutions is a
result of nontrivial structure of the Higgs vacuum 
${\cal{N}}=\{\phi^a: \phi^a \phi^a = v^2\}$
which has topology of a two sphere. The asymptotic Higgs fields 
$\phi_{r \rightarrow \infty}$ 
which are mappings from spatial infinity to the vacuum manifold of the Higgs
fields could be divided into infinite number of inequivalent homotopy
classes $\Pi_2({\cal{N}}) = Z$. The transition (by continuos deformation)
from one to another class needs infinite amount of the energy what
makes these solutions stable.

A particular example of the monopole is given by Polyakov - 't Hooft (P-T)
ansatz
\be
\phi^a = x^a H(r) \equiv \Phi^a , ~~~ {A_k}^a = \epsilon_{aki} x^i K(r) \equiv {W_k}^a , ~~~ {A_0}^a = 0,
\ee
where $H$ and $K$ have the following asymptotic behaviour
$H \stackrel{r \rightarrow 0}\longrightarrow H_* = const$, 
$K \stackrel{r \rightarrow 0}\longrightarrow K_* = const$ and 
$H \stackrel{r \rightarrow \infty}\longrightarrow {1 \over r} v$, 
$K \stackrel{r \rightarrow \infty}\longrightarrow - {1 \over e} {1 \over r^2}$.
Monopole of this form carries one unit of topological charge. 

From now we shall confine ourselves to $n=1$ monopole sector
of the Georgi - Glashow model.

Following the approach of paper \cite{Ost} we consider time deformations of the
P-T ansatz
\be
\phi^a = \Phi^a f(t) , ~~~ {A_k}^a = {W_k}^a g(t) , ~~~ {A_0}^a = 0,
\ee
where $f>0$ and $g>0$ are unknown functions of time.

Lagrangian density on this new ansatz takes the form
\be
{\cal L} = {1 \over 2} \left( \Phi^a \Phi^a \dot{f}^2 + {W_k}^a {W_k}^a \dot{g}^2 \right)
 - V(f,g)
\ee
where
\begin{eqnarray}
V & = &
 {1 \over 4} \lambda (\Phi^a \Phi^a f^2 - v^2)^2
+ {1 \over 4} {F_{sk}}^a {F_{sk}}^a g^2 
- {1 \over 2} e \epsilon^{abc} {F_{sk}}^a {W_s}^b {W_k}^c g^3
\nonumber \\
 & & 
+ {1 \over 4} e^2 [
({W_s}^a {W_s}^a)({W_k}^b {W_k}^b) 
- ({W_s}^a {W_k}^a)({W_s}^b {W_k}^b)] g^4
\nonumber \\
 & & + {1 \over 2} (\partial_k \Phi^a)(\partial_k \Phi^a) f^2
- e \epsilon^{abc} (\partial_k \Phi^a) {W_k}^b \Phi^c g f^2
\nonumber \\
 & & + {1 \over 2} e^2 [ ({W_k}^a {W_k}^a) (\Phi^b \Phi^b) -
({W_k}^a \Phi^a) ({W_k}^b \Phi^b)] g^2 f^2
\end{eqnarray}
As the homogenous time dependent shift of the fields (7) in the whole 
space costs infinite amount of the energy it is unphysical.
Therefore we confine our interest only to excitations of the 
core of the P-T monopole i.e.  located in region surrounding
zero of the Higgs fields.

We implement locality of these excitations by taking zero
($r \rightarrow 0$) asymptotics of the shape functions
$H$ and $K$
\be
{\cal L} = {1 \over 2} \left( r^2 {H_*}^2 \dot{f}^2 + 2 r^2 {K_*}^2 \dot{g}^2 \right)
 - V(f,g) ,
\ee
where
\begin{eqnarray}
V & = &
6 {K_*}^2 g^2 + 2 e r^2 {K_*}^3 g^3 + {1 \over 2} e^2 r^4 {K_*}^4 g^4 + {3 \over 2} {H_*}^2 f^2 +
\nonumber \\
 & & 
+ 2 e r^2 {H_*}^2 {K_*} f^2 g + 
e^2 r^4 {K_*}^2 {H_*}^2 f^2 g^2 + {1 \over 4} \lambda (r^2 {H_*}^2 f^2 - v^2)^2  .
\end{eqnarray}
The effective lagrangian of the spherical excitations is in natural way
defined by integration of the lagrangian density over a sphere of 
radius ${r_0}$ which contains the whole excitation
$L= \int_{S_{r_0}} d^3 x {\cal{L}} = 4 \pi \int_0^{r_0} d r r^2 {\cal{L}}$.
Local (regular in time) excitations have been found for large amount of
solitonic solutions \cite{Dobro}. We attempt to check possibility
of existence not only  regular but also chaotic solutions.
After rescaling by the overall factor ${4 \pi \over 5} {r_0}^5$ the monopole effective lagrangian 
$q^1 = f$, $q^2 = g$ has the form
\be
L = {1 \over 2} \left[ {H_*}^2 (\dot{q}^1)^2 + 2 {K_*}^2 (\dot{q}^2)^2 \right] - V(q^1,q^2) , 
\ee
\begin{eqnarray}
V & = &
{10 \over {r_0}^2} {K_*}^2 (q^2)^2 + 
2 e {K_*}^3 (q^2)^3 + {5 \over 14} e^2 {r_0}^2 {K_*}^4 (q^2)^4 
+ {5 \over 2 {r_0}^2} {H_*}^2 (q^1)^2 +
\nonumber \\
 & & 
+ 2 e {H_*}^2 {K_*} (q^1)^2 (q^2) + 
{5 \over 7} e^2 {r_0}^2 {K_*}^2 {H_*}^2 (q^1)^2 (q^2)^2 + 
\nonumber \\
 & &
+ {1 \over 4} \lambda \left( {5 \over 7} {r_0}^2 {H_*}^4 (q^1)^4 - 2 v^2 {H_*}^2 (q^1)^2
+ {5 \over 3 {r_0}^2} v^4 \right)
\end{eqnarray}
Now we shall use, based on the geodesic deviation equation,
local criterion of chaos.  In two dimensions 
system allows chaotic behaviour if the Gauss curvature
of the geometry determined by corresponding Jacobi metric
is negative.
If the Gauss curvature is positive and 
admissible for the physical trajectories
region has no boundary then chaos does not occur.
According to considerations of appendix (A8) the extrinsic 
curvature is 
\be
\hat{K} = {1 \over 4 (E - V)^3}
\left\{ (E - V) 
\left[ {1 \over {H_*}^2} {\partial^2 V \over \partial (q^1)^2 }
+ {1 \over 2 {K_*}^2} {\partial^2 V \over \partial (q^2)^2 } \right]
+ \left[ {1 \over {H_*}^2} \left({\partial V \over \partial q^1}\right)^2
+ {1 \over 2 {K_*}^2} \left({\partial V \over \partial q^2}\right)^2 \right]
\right\}
\ee
In admissible for trajectories region of the configuration space
$(E - V) > 0$.
The second square bracket is also positive.
Explicit differentiation of the potential (14) shows that 
${\partial^2 V \over \partial (q^2)^2 } > 0$
and
${\partial^2 V \over \partial (q^1)^2 } = - {1 \over 2} M_H + P(q^1,q^2)$,
where
$M_H = 2 \lambda v^2$
is a Higgs mass and
$P = {5 \over {r_0}^2} {H_*}^2 + 4 e {H_*}^2 {K_*} (q^2) + {10 \over 7} e^2 {r_0}^2 {K_*}^2 {H_*}^2 (q^2)^2
+ {15 \over 7} \lambda {r_0}^2 {H_*}^4 (q^1)^2 > 0$ is
positive function.
From equation (15) it follows that the sign of the Gauss curvature
is controlled by the value of the Higgs mass. 
For small Higgs masses $\hat{K}$ is positive and excitations
evolve in regular way.

If mass of  Higgs is sufficiently large then $\hat{K}$ became 
negative and chaos may occur.

\section{Remarks}
We considered spherically symmetric time - dependent deformations
of the P-T monopole. This deformation could be obtained
by the choice of the initial configuration which is deformed in the
center and pure P-T monopole (7) at spatial infinity.
In the heavy Higgs sector the local GDE criterion of chaos
reveals chaotic behaviour of the  fields. 
Because of the spherical symmetry we do not expect any 
radiation in this system which convince us that the
excitation does not disappear.

\hfill\break


\section{Appendix}
Our purpose is to calculate the Gauss curvature $\hat{K}$ of the
surface described by the Jacobi metric. Let us start from evaluating
the components of the Riemann tensor. The most efficient way of 
evaluating the components of the curvature tensor has its origin
in Cartan structural equations
$$
d \Theta^a + {{\omega}^a}_b \wedge \Theta^b = 0 ,
\eqno({\rm A}.1)
$$
$$
{\hat{R}^a}_b = d {{\omega}^a}_b + {{\omega}^a}_c \wedge {{\omega}^c}_b ,
\eqno({\rm A}.2)
$$
where $\Theta^a$ is the orthogonal co-base, i.e. such that 
$ \hat{g} = \Theta^1 \otimes \Theta^1 + ... + \Theta^N \otimes \Theta^N . $
The first equation ensures the torsionless of the connection 
and the second one is a definition of the curvature 2 - form.

Let us confine to the two dimensional manifold equipped 
in metric 
$$
\hat{g} = \alpha dx \otimes dx + \beta dy \otimes dy ,
\eqno({\rm A}.3)
$$
where $\alpha = \alpha(q^1,q^2)$, $\beta = \beta(q^1,q^2)$.

In orthonormal co-base  
$\Theta^1 = \sqrt{\alpha} d q^1$,
$\Theta^2 = \sqrt{\beta} d q^2$ metric
(A3) can be expressed in standard form
$$
\hat{g} = \Theta^1 \otimes \Theta^1 + \Theta^2 \otimes \Theta^2 .
\eqno({\rm A}.4)
$$
In two dimensions the definition of the curvature 2-form
(A2) is even simpler
$$
{\hat{R}^a}_b = d {\omega^a}_b .
\eqno({\rm A}.5)
$$
Evaluation of the curvature is extremely simple because the 
connection 1-form has the only one independent component
(which follows from equation $d \hat{g} = 0$)
$ {\omega^1}_2 = - {\omega^2}_1 = \omega_{12} = - \omega_{21} $.
The first Cartan structural equation leads to
$$
{\omega^1}_2 = {1 \over 2 \sqrt{\alpha \beta}} \left(
{1 \over \sqrt{\alpha}} {\partial \alpha \over \partial q^2} \Theta^1
- {1 \over \sqrt{\beta}} {\partial \beta \over \partial q^1} \Theta^2
 \right) .
\eqno({\rm A}.6)
$$

On the other hand we use the fact that ${\hat{R}^1}_2$
component of the Riemann curvature tensor can be
expressed by Gauss curvature
$$
d {\omega^1}_2 = {\hat{R}^1}_2 = \hat{K} \Theta^1 \wedge \Theta^2 .
\eqno({\rm A}.7)
$$
Now the Gauss curvature computation is a straightforward
substitution of ${\omega^1}_2$ into equation (A7)
$$
\hat{K} = - {1 \over 2 \alpha \beta} \left( 
{\partial^2 \alpha \over \partial {(q^2)}^2}
+ {\partial^2 \beta \over \partial {(q^1)}^2}
\right)
+ {1 \over 4 \alpha^2 \beta^2} \left[
\left( {\partial \alpha \over \partial q^2} \right) 
{\partial \over \partial q^2} (\alpha \beta) +
\left( {\partial \beta \over \partial q^1} \right) 
{\partial \over \partial q^1} (\alpha \beta) \right] .
\eqno({\rm A}.8)
$$

\eject


\begin{thebibliography}{99}
\bibitem{Gucken} 
J.Guckenheimer, P.Holmes, "Nonlinear Oscilations, 
Dynamical Systems and Bifurcations of Vector Fields," 
Springer Verlag, 1983.\\ \
L.E.Reichl, "The Transition to Chaos,"
Springer Verlag, 1992.\\ \
\bibitem{Ott}
E.Ott, "Chaos in Dynamical Systems,"
Cambridge University Press, 1993. \\ \
\bibitem{Perko}
L.Perko, "Differential Equations and Dynamical Systems,"
Springer Verlag, 1992.\\ \
\bibitem{Toda}
M.Toda, Phys. Lett. {\bf 48 A} (1974) 5. \\ \
\bibitem{Arnold}
V.I.Arnold, "Mathematical Methods of Classical Mechanics,"
Springer Verlag, New York 1978.\\ \
M.Szyd\l{}owski, J.Szcz\c{e}sny, Phys. Rev. D {\bf 50} (1994) 819.\\ \
\bibitem{Tze}
I.Bars, A.Chodos, C.H.Tze, "Symmetries in Particle Physics,"
Plenum Press, New York and London 1984.\\ \
\bibitem{Gal}
E.E.Donets, Preprint gr-qc 9807057 \\ \
\bibitem{Zhang}
S.C.Zhang, Science {\bf 275} (1997) 1098.\\ \
E.Demler, S.C.Zhang, Preprint cond-mat 9805404.\\ \
\bibitem{Dobro}
H.Arod\'z, Nucl.Phys. B {\bf 509} (1998) 273.\\ \
H.Arod\'z, L.Hadasz, Phys.Rev. D {\bf55} (1997) 942.\\ \
E.Braaten, S.Townsend, L.Carson, Phys.Lett. B {\bf 235} (1990) 147.\\ \
T.Dobrowolski, Phys.Rev. D {\bf 50} (1994) 6503. \\ \
C.Houghton, N.Manton, P.Sutcliffe, Nucl.Phys. B {\bf 510} (1998) 507.\\ \
\bibitem{Ost}
L.Salasnich, Preprint nucl-th 9707035.\\ \
\end{thebibliography}
\end{document}